\documentstyle[aps,pra,psfig,epsf,epsfig,twocolumn]{revtex}

\begin{document}

%\draft
\title{
On the possibility of quantum computation based on photon 
exchange interactions
}
\author{T. Opatrn\'{y}$^{1,2}$ and G. Kurizki$^{3}$
 }
\address{
$^{1}$ 
Theoretisch-Physikalisches Institut, Friedrich-Schiller Universit\"{a}t,
Max-Wien-Platz 1, D-07743 Jena, Germany \\
$^{2}$ 
Department of Theoretical
Physics, Palack\'{y} University, Svobody 26, CZ-77146 Olomouc, Czechia \\
$^{3}$ 
Department of Chemical Physics, Weizmann Institute of Science,
761~00 Rehovot, Israel
}
\date{\today}

\maketitle

\begin{abstract}

We examine several proposed schemes by Franson et al.  for quantum logic gates
based on non-local exchange interactions between two photons in a medium.   In
these schemes the presence of a {\em single} photon in a given mode is supposed
to induce a large phase shift on another photon  propagating in the same
medium. We conclude that the schemes proposed so far are not able to produce
the required conditional
phase shift, even though the proposals contain many 
stimulating and intriguing ideas.

\end{abstract}

%%%%%%%%%%%%%%%%%%%%%%%%%%%%%%%%%%%%%%%%%%%%%%%%%%%%%%%%

\section{Introduction}

The fact that photons almost do not interact with each other limits our
ability to build  photonic logical gates for two and more qubits.  Nonlinear
effects whereby one light beam influences another require large numbers of
photons or else photon confinement in high-$Q$ cavities \cite{Brune,kim}.
Therefore a gate in which one photon-qubit would influence another is
difficult to construct. This difficulty has motivated Franson et al.
\cite{Franson,FP98,FP99,FPD99} to search for a new effect that would enable
the creation of a simple gate operating at the two-photon level.

Their basic proposal for the photonic computer is introduced in \cite{FP98}.
Each qubit is represented by a photon which can travel along two alternate
paths in an interferometer. Single qubit manipulation can be easily performed
by setting proper phases in the relevant interferometer. The reading of the 
computed result is done by photodetection.  The initial state should be
prepared as a multi-mode product of single-photon Fock states. The authors of 
\cite{FP98} propose using quantum non-demolition photon number measurements to
select such states from a larger set of coherent states. 

The essential and most difficult part of the proposed scheme is an efficient
two-bit quantum gate. Physically, the phase which one photon picks up in an
interferometer should be determined by the path which the other photon chooses
in another interferometer. For this to occur both interferometers should share
one common branch in a nonlinear medium, such that when  both photons travel
along this branch an extra phase equal to $\pi$ would arise. This would require
a giant cross-Kerr effect with  negligible photon loss. The main advantage of
such a scheme would be experimental simplicity, as compared to cavity-based
schemes \cite{Brune,kim}. 

Here we examine the main assumptions and results of the schemes
detailed in Ref. \cite{Franson,FP98,FP99,FPD99}. 
We concentrate on the physical principle for the logic gates, leaving
out the questions of the preparation and detection stages. 
In 
Sec.~\ref{nonloc} an overview of these schemes
is outlined. In Sec.~\ref{critics} we  present our detailed 
evaluation. Our conclusions
are given in Sec.~\ref{conclusion}.

%%%%%%%%%%%%%%%%%%%%%%%%%%%%%%%%%%%%%%%%%%%%%%%%%%%%%%%%

\section{Overview of non-local photon exchange schemes}
\label{nonloc}

The underlying model of \cite{Franson,FP98} can be described as follows. Two
light pulses simultaneously enter a medium of $N$ off-resonant atoms. The two
pulses contain $n_{1,2}$ photons, respectively.  Applying perturbation theory
we can find the energy eigenvalues of the joint system of atoms and photons 
(within the dipole and rotating-wave approximation).  Some of the fourth-order
perturbation terms contain products of the photon numbers $n_1n_2$.  Such
terms can be characterized by Feynman diagrams in which a photon from pulse 1
is virtually absorbed by atom A and re-emitted by atom B, whereas a  photon
from pulse 2 is absorbed by atom B and re-emitted by atom A. The number of
such terms is proportional to the number of various atomic pairs, i.e.,
$N(N-1)/2$. If such terms contributed to the total energy, the physical result
would be a nonlinear refractive index of the medium with very interesting
properties: (i) the non-linear term would be proportional to the number of
atoms squared $N^2$; (ii) the index of refraction caused by pulse 1 would be
proportional to the photon number in pulse 2 and vice versa. However, after
summing them up, all the terms containing  $n_1n_2$ exactly cancel each other.
We may ask whether such non-linear terms are just a mathematical artifact of
perturbation theory or whether they correspond to some real physical
situation. If the latter is true, the question is how to suppress some of the
terms so that  the remaining terms contribute to an experimentally observable
non-linear effect.

Franson's first suggestion for suppressing some of the  $n_1n_2$ terms was to
take advantage of collisional line broadening \cite{Franson}.  The model used
$N$ two-level atoms and  the resulting non-linear part of the total atom-photon
energy was claimed to be
\begin{eqnarray}
 \Delta E \approx - \frac{2M^4N^2 n_1n_2 f_R}{\delta ^3}
 \frac{w^2}{(\delta_1-\delta_2)^2},
 \label{rce1}
\end{eqnarray} 
where $M$ is the transition matrix element, $f_R$ ($< 1$) is a factor taking
into account decoherence due to a possible which-way information about the
position of the  photon absorption and re-emission, $\delta_{1,2}$ $\approx$ 
$\delta$ are the detunings of the modes 1 and 2 from the atomic resonance, and
$w$ is the collisional line-width. For an efficient non-linear 
coupling between
single-photon pulses one should find a sufficiently dense medium ($N$ large) and a
broad collisional line-width, so that $w^2/(\delta_1-\delta_2)^2$ $\approx 1$,
which would yield a considerable non-linear energy shift.

The next suggestion was to manipulate the atomic resonance frequencies
\cite{FP99,FPD99}. The authors considered three-level atomic media, where
strong laser pulses coupled to one of the atomic transitions would 
manipulate the resonance frequency of another transition, e.g., by 
Stark shifts. Thus, the medium would be turned on and off resonance with the
incident photons, which would be absorbed and re-emitted in a controlled way.

In Ref. \cite{FPD99} it is assumed that the photonic states would be coupled to
collective $N$-atom excitations (Dicke states). The coupling  frequency is then
proportional to $\sqrt{N}$ and is assumed to be  larger than any decay and
decoherence rates.    In both  schemes with external driving pulses the authors
claim that the photon of one kind would exhibit Rabi oscillations whose
frequency depends on the presence or absence of the photon of the other kind. A
proper choice of the external strong laser pulses is then claimed to
effectively induce the required non-linear coupling of  two single-photon
pulses.

%%%%%%%%%%%%%%%%%%%%%%%%%%%%%%%%%%%%%%%%%%%%%%%%%%%%%%%%

\section{Detailed examination of the schemes}
\label{critics}

%%%%%%%%%%%%%%%%%%

\subsection{Collisional scheme}

In the derivation of the results of \cite{Franson} there are several
unjustified
assumptions: 

(1) The model of   \cite{Franson} assumes  two weak off-resonant light pulses
propagating in a medium of two-level atoms. It is supposed that the effective
Hilbert space describing the system is spanned by quantum states with
different numbers of photons in the two relevant optical modes (1 and
2) and with
different excited and de-excited atoms. All other optical modes are ignored:
processes where photons can be re-emitted to modes other than 1 and 2 are
disregarded. This is an arbitrary assumption: in open space photons
are re-emitted into a continuum of modes, so that
the main effect would be scattering. This would, of course, invalidate the
potential application of the proposed effect as a quantum gate.

(2) Even if we go along with the model where only two optical modes are present, we
cannot accept the main result Eq. (\ref{rce1}). Its derivation is based on
replacing in the fourth-order perturbation expansion the energy levels
$\epsilon_m$ which are influenced by collisions by the complex values 
$\epsilon_m$ $-$ $iw$. However, in doing so, we always obtain {\em 
zero} for the
non-linear $n_1 n_2$ terms. The only way to obtain Eq.~(\ref{rce1}) is
to assume that the states with {\em no excited atoms\/} ($n_1\pm1$
photons in mode 1  and $n_2\mp1$ in mode 2) suffer from collisional
decoherence and their energy levels should be modified by adding the $iw$
term. Of course, this assumption is not physically justified.

(3) Finally, even under the unlikely assumption that {\em states with all atoms in
 the ground
level decohere by collisions}, the non-linear term (\ref{rce1}) would be
accompanied by an imaginary part
\begin{eqnarray}
   \Delta E ' = - 2 i f_{R} M^4 N^2 n_1 n_2 \frac{w}{\delta^2 (\delta_1 -
   \delta_2)^2} ,
\end{eqnarray}
whose magnitude is {\em larger} by a factor of $\delta/w$ than the real part.
Hence,  decay would always dominate any such non-linear phase shift
and render the effect unobservable.
We note that in Ref. \cite{FP99} the authors
mention that there are difficulties with the collisional scheme.

%%%%%%%%%%%%%%%%%%

\subsection{Laser-induced nonlinear phase shift in ``ladder'' systems}

In the scheme of \cite{FP99}  one applies strong laser pulses which
induce AC Stark shifts and thereby
change the detuning of the near-resonant atomic transition from the relevant
single-photon-carrying modes. Again, this model  assumes that after photons 1
and 2 are absorbed by the atoms (not only virtually but also really, when the
atoms are on resonance), they can only be re-emitted into the useful modes 1
and 2.  Thus, it is assumed in  \cite{FP99} that in {\em open space},  a
single optical photon on resonance with the atomic medium can {\em perform a
Rabi oscillation without being scattered to the continuum of other modes}.  In
other words,  a resonant atom which absorbs the only photon  from the
traveling field would re-emit it to  {\em exactly the same\/}  (now empty)
mode. In Ref. \cite{FP99}, no mechanism has been presented that would
justify the assumed mode selectivity.

On the other hand, if the mode selectivity is guaranteed (e.g., by a cavity
or by the Dicke cooperativity mechanism described in \cite{FPD99}, see below), then 
for off resonant photons a weak non-linear phase shift may occur, whose magnitude
is of the same order as their absorption probability. 
We think that a two-photon interference 
experiment revealing such  
a phase shift would be an interesting (though demanding) challenge.
We feel, however, that such a phase shift would be too small to be useful
for quantum logic gates.

%%%%%%%%%%%%%%%%%%

\subsection{The Dicke cooperative mechanism in Raman transitions}

A mechanism supporting mode selectivity and the elimination of the mode
continuum is presented in the e-print \cite{FPD99}. It is argued there that a
field mode of the wavevector ${\bf k}$ is effectively only coupled to a
particular superposition of atomic excited states (Dicke state \cite{Dicke}),
namely 
\begin{eqnarray} 
 | p ({\bf k}) \rangle = \frac{1}{\sqrt{N}} 
 \sum_j \exp (i {\bf k}\cdot {\bf r}_j) |e_j
 \rangle , 
\label{3}
\end{eqnarray} 
where  $|e_j \rangle$ determines the state with $j$th
atom excited and  all the other atoms being in the ground state; the summation
runs over all $N$ atoms. The coupling between the field mode and the
corresponding Dicke state is $\sqrt{N}M$, $M$ being the coupling between
the field 
and a single atom. In an original extension of the standard formalism
in \cite{FPD99}
the same cooperative enhancement is shown to apply to Raman transitions:
The Dicke collective state will then be excited for the ground state
by the operator
\begin{eqnarray}
\hat{R}_+(\vec{k}-\vec{K})=\sum_{j=1}^N\hat{R}_+^{(j)}
e^{(\vec{k}-\vec{K})\cdot\vec{r}_j}
\end{eqnarray} 
where $\vec{k}$ and $\vec{K}$ are the wavevectors of the incident
photon and the external laser field, respectively, and
$\hat{R}_+^{(j)}$ is the $j$th atom raising operator. 

The spectacular
feature of the Dicke formalism is that, in the absence of decoherence
or losses, {\em any distribution} of atomic positions is guaranteed to
have a state of maximal cooperation, such that the corresponding
transition rate (Rabi frequency) is enhanced by $\sqrt{N}$ compared
to that of a single atom. For Raman transitions we estimate that the
cooperatively enhanced Rabi frequency is (in SI units)
\begin{eqnarray}
\Omega^{\rm (coop)}_{\rm Raman}\approx
\sqrt{\rho\omega/(\epsilon_0\hbar)}\mu \Omega_0 /\Delta
\label{5}
\end{eqnarray} 
where $\rho$ is the atomic  density, $\omega$ the transition  frequency,
$\epsilon_0$ the vacuum permittivity, $\mu$ the transition dipole moment,
$\Delta$ the detuning from the single-photon resonance and 
$\Omega_0 =\mu E/\hbar$ is
the strong-field Rabi frequency. 

The crux of the effect is that during the relevant time, energy can only be
exchanged between two states - a photon in a single field mode and a
single-excitation  atomic collective state (\ref{3}),  thus exhibiting Rabi
oscillations.  The decay and decoherence rates leading to scattering into other
modes  could be presumably slower, due to the $\sqrt{N}$ proportionality of the
Rabi frequency. Because of linearity and the weak dependence of the Rabi
frequency on the photonic frequency $\omega$, the same would be valid for
photonic wavepackets. Under these assumptions, any sufficiently dense  medium
of resonant (or Raman-resonant) atoms could be transparent for single-photon
light pulses: the photon would travel inside the medium ``dressed" by the atomic
excitations. 

%%%%%%%%%%%%%%%%%%

\subsubsection*{Evaluation} 

The cooperative Dicke effect discussed in  
\cite{FPD99} is essentially the well-known excitonic enhancement of
absorption and emission in crystals \cite{dav62} except that Raman
transitions are discussed in \cite{FPD99} instead of the standard
direct transitions. 
Thus far, single-photon Rabi oscillations associated with cooperative
(excitonic) enhancement have only been observed in semiconducting {\em
cavities} \cite{rar96}, where they have the character of the single-mode
Tavis-Cummings\cite{har} cooperative effect known for atoms in high-$Q$ 
cavities \cite{Raizen}. 
By contrast, single-photon cooperative Rabi
oscillations {\em in open space} of mode continuum suggested in 
\cite{FPD99} have never been observed. The reason is that decoherence
usually prevails, i.e., is {\em faster} than the achievable Rabi
oscillation. This can be clarified using the estimate (\ref{5}) in the
two possible regimes:

(a) In the {\em high-density} regime, $\rho/k^3\gg 1$, corresponding to
interatomic distances $r_{ij}\sim\rho^{-1/3}$ much smaller than the photon
wavelength, the dominant source of decoherence detrimental to cooperation are
resonant dipole-dipole interactions \cite{cra} whose rate is $\Omega_{\rm
dip}\sim\gamma/(kr_{ij})^3$, $\gamma$ being the radiative linewidth (for direct
transitions), i.e.,  $\Omega_{\rm dip}$ scales as $\rho/k^3$. For both direct
transitions and Raman transitions Eq. (\ref{5}) typically yields a lower rate
than the dipole-dipole rate. Only for spatially {\em symmetric} atomic
arrangements cooperative effects prevail over the dipole-dipole dephasing
\cite{kur96}. For example, for interatomic distances of 10 nm $\rho\sim
10^{18}$ cm$^{-3}$,  $\Omega^{\rm (coop)}_{\rm Raman}\alt 10^{12}$
s$^{-1}\alt\Omega_{\rm dip} \sim 10^6\gamma$.

(b) In the {\em low-density} regime $\rho/k^3\alt 1$, the dipole-dipole rate is
less than $\gamma$ and does not have to hamper cooperation. However, other
sources of dephasing set $T_2$ (the decoherence time during which
the off-diagonal density matrix elements go to zero)
to be shorter than the cooperative Rabi period
(typically  $ \Omega^{\rm (coop)}_{\rm Raman}\alt 10^{6}$ s$^{-1}$) both in
thermal gases ($T_2\alt 10^{-10}$ s)  and in semiconductors  ($T_2\alt
10^{-12}$ s). 

However, the most important, ingenious assumption in \cite{FPD99} concerns the
initial  condition for the atom-field system: the atoms are suddenly switched
on resonance in the vicinity of the photon  wavepacket, which is already within
the medium. This is achieved by an appropriate geometry in which the photonic 
wavepacket overlaps with the strong rapidly-switched laser pulse, propagating
perpendicularly to the single photon.  
Thus, only a
photon being initially within the resonant medium can perform Rabi
oscillations with the corresponding Dicke state. This is in contrast with the
usual situation when a photon arrives at the medium which already has been
resonant: the photon is then reflected or absorbed at the medium boundary, but
cannot enter inside.
We think that if this intriguing effect is realizable, its
experimental observation  would be very 
interesting. The following experiment could be planned:
Photons would be sent into a transparent medium
one by one.
When inside the medium, a strong perpendicularly propagating pulse would bring the 
medium into resonance with the photons. The probability of detecting the photons 
as they exit the medium would 
be a periodic function of the duration of the strong pulse, thus demonstrating the
single-photon Rabi oscillations.

Let us mention that several mechanisms of mapping the quantum state of light 
onto a collective state of atoms in the presence of a strong field 
have been suggested recently \cite{EIT}. 
The challenge of such mechanisms would be to allow coherent excitation 
and de-excitation of collective states with {\em single\/} photons.
A detailed comparison of these intriguing
approaches would be of great interest.

%%%%%%%%%%%%%%%%%%%%%%

\subsection{Two-photon entanglement and conditional phase shifts}

Notwithstanding the chances of realizing the cooperative effect discussed
above,  the question is whether this effect can be used to produce the required
conditional phase shift of the photonic states. In \cite{FP99,FPD99} the
authors argue that a proper sequence of external laser pulses  would accomplish
this task. Concrete suggestions for the pulse sequences were given in 
\cite{FP99} (five-pulse sequence), and in \cite{FPD99} (three-pulse sequence).
In the following we discuss these two suggestions separately.

%%%%%%%%%%%

\subsubsection{Five-pulse sequence}

The first pulse brings the medium into resonance with photon 1 causing a $\pi$
Rabi transition: if photon 1 was initially present, it is absorbed creating a
single-excitation Dicke state. The second pulse brings the medium into
resonance with photon 2. The Rabi frequency now depends on whether the medium
is excited or not. The authors of \cite{FP99} assume that the second pulse
causes a $2\pi$ transition if there is no initial excitation (photon 1 is
absent) or a  $\sqrt{2}2\pi$ transition if there is initial medium excitation
(photon 1 is present). In the latter case it is assumed that a superposition of
two states is produced: a state with two  atomic excitations and no photon, and
a state with a single atomic excitation and a single photon in mode 2. The
third pulse is used to produce a phase-shift in this superposition, and the
remaining two pulses reverse the evolution of the first two pulses, to a state
with the initial number of photons.

Our objection is that during the second pulse, also a state with two photons in
mode 2 can be produced (as can be seen from the Hamiltonian, Eq. (31) of
\cite{FP99}): stimulated photon emission would occur with the same rate as
photon absorption. Of course, the presence of such a two-photon state would
invalidate the function of a quantum gate, where qubits are represented by
single-photon states.

%%%%%%%%%%%

\subsubsection{Three-pulse sequence}

The above flaw is removed in the scheme of \cite{FPD99}: during the second
pulse in the case of initially one atomic excitation and one photon in mode 2,
the system oscillates between this state and the superposition of the state
with two atomic excitations and no photons and the state with two photons and
no atomic excitations. The frequency of this oscillation is twice as large as
the Rabi frequency in the absence of the initial atomic excitation. Thus, if
there was no photon in mode 1, the photon 2 exhibits a $2\pi$ Rabi transition
during the second pulse, whereas if there was a photon in mode 1, the system
exhibits a $4\pi$ Rabi transition. A  $2\pi$ Rabi transition returns the
original state with additional sign $-1$, whereas a  $4\pi$ Rabi transition
simply reproduces the original state. The authors of  \cite{FPD99} claim that
this difference in sign produces the required conditional phase shift.

Let us investigate in more detail the evolution of the states, denoting the 
basis of the logical gate as $|0,0\rangle$,  $|0,1\rangle$,  $|1,0\rangle$,
and   $|1,1\rangle$. Here, e.g., $|1,0\rangle$ means that photon 1 goes through
the medium whereas photon 2 does not, etc. The evolution is then as follows.
(a) State  $|0,0\rangle$: there are no Rabi transitions, therefore
$|0,0\rangle$ $\to$ $|0,0\rangle$. (b) State    $|0,1\rangle$: no change during
the first pulse (photon 1 is absent),  a $2\pi$ Rabi transition during the
second pulse and no change during the third pulse, therefore   $|0,1\rangle$ 
$\to$  $-|0,1\rangle$. (c) State  $|1,0\rangle$: a $\pi$ transition during the
first pulse,  a $2\pi$ transition during the second pulse and  a $\pi$
transition during the third pulse, therefore $|1,0\rangle$  $\to$ 
$|1,0\rangle$. (d) State   $|1,1\rangle$: a   $\pi$ transition during the first
pulse, a  $4\pi$ transition during the second pulse and a $\pi$ transition 
during the third pulse, therefore  $|1,1\rangle$  $\to$  $-|1,1\rangle$.  Thus,
the transformation matrix is
\begin{eqnarray}
U = \left(   
\begin{array}{rrrr}
1 & 0 & 0 & 0 \\
0 & -1 & 0 & 0 \\
0 & 0 & 1 & 0 \\
0 & 0 & 0 & -1
\end{array}
\right) ,
\end{eqnarray}
which is {\em not\/} the required conditional phase shift. This  transformation
{\em does not} entangle the two photons and is thus not suitable for building
quantum logical gates.

%%%%%%%%%%%%%%%%%%%%%%%%%%%%%%%%%%%%%%%%%%%%%%%%%%%%%%%%%%%%%%%%%

\section{Conclusion}
\label{conclusion}

In the works \cite{Franson,FP98,FP99,FPD99} we have not found a convincing
proof that the suggested mechanisms could produce the  conditional phase shift
required for a quantum gate. On the other hand, we cannot disprove
such a mechanism altogether, i.e., claim that it is principally impossible.
Be it as it may, 
we find the idea of coupling the photonic state to  an atomic Dicke state by
fast switching,   which would result in single-photon Rabi oscillations, very
interesting and ingenious, even though the prospects for its realization are
presently unclear.

{\bf Note added in proof:} A proof has been given \cite{Mischa} that
exchange interactions between photons in a large ensemble of atoms cannot
generate entanglement: if the process generates two photons in
distinguishable modes, then the two-mode state can be factorized.

\section*{Acknowledgments}

We thank 
A. Ben-Reuven, 
I. Cirac,
M. Fleischhauer,
Ph. Grangier,
A. Kofman, 
A. Kozhekin,
M. Lukin,
J. Pe\v{r}ina,
S. Scheel,
E. Schmidt,
Y. Silberberg 
and D.-G. Welsch
for stimulating discussions.
This work was supported by ISF
and DFG grants.

\end{document}